\documentclass[12pt,epsfig]{article}

\usepackage{amssymb,epsfig}

\usepackage{psfrag}

\psfrag{kt}{{\footnotesize $k_\perp$}}
\psfrag{=}{{\footnotesize $=$}}
\psfrag{+}{{\footnotesize $+$}}
\psfrag{l}{{\footnotesize $l$}}
\psfrag{p1z'}{{\footnotesize $p_1z'$}}
\psfrag{q1}{{\footnotesize $q_1$}}
\psfrag{p1}{{\footnotesize $p_1$}}
\psfrag{-p1bz'}{{\footnotesize $-p_1{\bar z}'$}}
\psfrag{-q2}{{\footnotesize $-q_2$}}
\psfrag{x1p2}{{\footnotesize $x_1p_2$}}
\psfrag{x2p2}{{\footnotesize $x_2p_2$}}
\psfrag{p2}{{\footnotesize $p_2$}}
\psfrag{p2'}{{\footnotesize $p_2'$}}
\psfrag{kpx1p2}{{\footnotesize $k_\perp, x_1$}}
\psfrag{kpx2p2}{{\footnotesize $k_\perp, x_2$}}
\psfrag{pi}{{\footnotesize $\pi$}}
\psfrag{a}{{\footnotesize $(a)$}}
\psfrag{b}{{\footnotesize $(b)$}}
\psfrag{c}{{\footnotesize $(c)$}}
\psfrag{d}{{\footnotesize $(d)$}}
\psfrag{e}{{\footnotesize $(e)$}}
\psfrag{f}{{\footnotesize $(f)$}}
\psfrag{g}{{\footnotesize $(g)$}}
\psfrag{Fipi}{{\footnotesize $\phi_\pi(z')$}}
\psfrag{Fchi}{{\footnotesize $F_\xi(x_1)$}}

\psfrag{I2}{{$|{\cal I}|^2$}}
\psfrag{z}{{$z$}}
\psfrag{TH}{\footnotesize $T_H(z',x_1)$}

\newcommand{\beq}[1]{
\begin{equation}\label{#1}}
\newcommand{\eeq}{\end{equation}}
\newcommand{\bea}[1]{
\begin{eqnarray}\label{#1}}
\newcommand{\eea}{\end{eqnarray}}
\newcommand\re[1]{(\ref{#1})}
\newcommand{\beqar}[1]{\begin{eqnarray}\label{#1}}
\newcommand{\eeqar}{\end{eqnarray}}
\begin{document}

\begin{titlepage}

\begin{flushright}
\begin{tabular}{l}
 TPR -- 01 -- 04\\
 hep-ph/0103275
\end{tabular}
\end{flushright}
\vspace{1.5cm}

\begin{center}
{\LARGE \bf
QCD factorization for the pion diffractive dissociation to two jets}
\vspace{1cm}

{\sc V.M.~Braun}${}^1$,
{\sc D.Yu.~Ivanov}${}^{1,2}$
{\sc A.~Sch\"afer}${}^1$ and
{\sc L.~Szymanowski}${}^{1,3}$
\\[0.5cm]
\vspace*{0.1cm} ${}^1${\it
   Institut f\"ur Theoretische Physik, Universit\"at
   Regensburg, \\ D-93040 Regensburg, Germany
                       } \\[0.2cm]
\vspace*{0.1cm} ${}^2$ {\it
Institute of Mathematics, 630090 Novosibirsk, Russia
                       } \\[0.2cm]
\vspace*{0.1cm} ${}^3$ {\it
 Soltan Institute for Nuclear Studies,
Hoza 69,\\ 00-681 Warsaw, Poland
                       } \\[1.0cm]

\vskip2cm
{\bf Abstract:\\[10pt]} \parbox[t]{\textwidth}{
  We calculate the cross section of a pion diffraction dissociation 
  in two jets with large transverse momenta originating from a hard 
  gluon exchange between the pion constituents.
  To the leading logarithmic accuracy (in energy),
  the contribution coming from small transverse separations
  between the quark and the antiquark in the pion 
  acquires the expected factorized form, the longitudinal momentum 
  distribution of the jets being proportional to the pion distribution
  amplitude. The hard gluon exchange can in this case be considered as a
  part of the unintegrated gluon distribution. Beyond the leading logarithms
  (in energy) this proportionality does not hold. Moreover, the collinear
  factorization appears to be broken by the end-point 
  singularities. Remarkably enough, the longitudinal momentum distribution
  of the jets for the non-factorizable contribution is calculable, and 
  turns out to be the same as for the factorizable contribution with 
  the asymptotic  pion distribution amplitude.
}
\vskip1cm
\end{center}

\vspace*{1cm}

\end{titlepage}

{\large \bf 1.~~} It has been conjectured a long time ago \cite{BBGG81}
that pion diffraction dissociation on a heavy nucleus $\pi A\to XA$
is sensitive to small transverse size configurations of pion constituents.
It was later argued \cite{FMS,NSS99} that selecting a specific hadronic 
final state that consists of a pair of (quark) jets with large transverse 
momentum $q_{1\perp}\simeq -q_{2\perp}$ one can obtain important 
insight into the pion 
structure as it turns out that the longitudinal momentum fraction 
distribution of the jets follows that of the pion valence parton 
constituents. A measurement of hard dijet coherent production on nuclei
presents, therefore, the exciting possibility of a direct measurement 
of the pion distribution amplitude and provides 
striking evidence \cite{E791} that this distribution is close to 
its asymptotic form. 

\begin{figure}[t]
\centerline{\epsfxsize7.0cm\epsffile{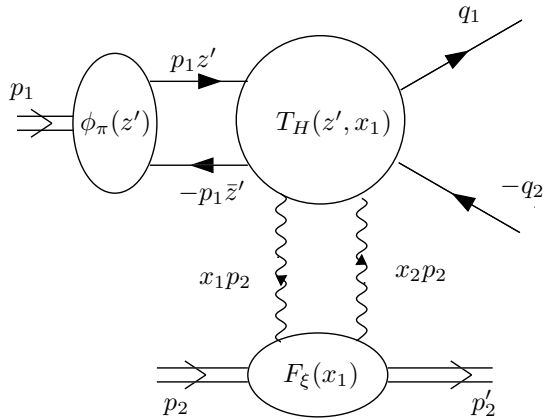}}
\caption[]{\small 
Kinematics of the coherent hard dijet production $\pi\to 2\, {\rm jets}$.  
The hard scattering amplitude $T_H$ contains at least one hard gluon 
exchange.
 }
\label{fig:1}
\end{figure}

From the theoretical point of view, the principal question is whether 
the relevant transverse size of the pion $r_\perp$ 
(alias the scale of the pion distribution amplitude $\mu = 1/r_{\perp}$) 
is determined by the color transparency condition 
$\mu \sim A^{1/3}\Lambda_{\rm QCD}$ or whether it is of the order of the 
transverse momenta of the jets $\mu \sim q_\perp$. In the latter case
one could envisage a factorization formula for the amplitude 
of hard dijet coherent production of the type (cf. Fig.~\ref{fig:1})
\beq{factor}
{\cal M}_{\pi \to 2\, {\rm jets}}
=\int\limits^1_0 dz^\prime \int\limits^1_0 dx_1 \,\phi_\pi
(z^\prime,\mu_F^2)\,T_H(z^\prime , x_1, \mu_F^2)\,F^g_\zeta(x_1, \mu_F^2)\,.
\eeq     
Here  $\phi_\pi (z^\prime,\mu_F^2)$ is the pion distribution amplitude,
and $F^g_\zeta(x_1, \mu_F^2)$ is the non-forward (skewed) gluon distribution 
\cite{Rad96,Ji97} in the target nucleon or nucleus; 
$x_1$ and $x_2=x_1-\zeta$ are the 
momentum fractions of the emitted and the absorbed gluons, respectively.
(The asymmetry parameter $\zeta$ is fixed by the process kinematics,
see below.) 
$T_H(z^\prime , x_1, \mu_F^2)$ is the hard scattering amplitude
and $\mu_F$ is the (collinear) factorization scale.

To the best of our knowledge, this question has never been studied.
As a first step in this direction, in this letter
we present an explicit calculation of the 
leading-order contribution to the imaginary part of 
$T_H(z^\prime , x_1, \mu_F^2)$ corresponding to a single hard gluon exchange.  
Since we are not interested in the color transparency phenomena,
we restrict ourselves to scattering from a single nucleon.
We will find that the structure of the hard gluon exchange is such that
it generates an enhancement by a logarithm of the energy in the region 
$z'\simeq z$ where $z$ is the energy fraction carried by the (quark) jet.
If only this logarithmic contribution is retained, the longitudinal 
momentum distribution of the jets indeed follows the shape
of the  pion distribution amplitude \cite{FMS,NSS99}. 
The hard gluon exchange can in this case be considered as a
part of the unintegrated gluon distribution, as advocated in \cite{NSS99}. 
 Beyond the leading logarithms
(in energy) this proportionality does no longer hold. Moreover, the collinear
 factorization \re{factor} appears to be broken by the end-point 
 singularities. Remarkably enough, the longitudinal momentum distribution
 of the jets for the non-factorizable contribution is calculable, and 
 turns out to be the same as for the factorizable contribution with 
 the asymptotic  pion distribution amplitude.

\vskip0.3cm
{\large \bf 2.~~} The kinematics of the process is shown in Fig.~\ref{fig:1}.
For definiteness, we consider pion scattering from a nucleon target.
%
%
The momenta of the incoming pion, incoming nucleon and the outgoing nucleon  
are $p_1, p_2$ and $p_2^\prime$, respectively. 
The pion and the nucleon masses are both neglected, 
$p_1^2=0$, $p_2^2=(p_2^\prime)^2=0$. 
We denote the momenta of the outgoing quark and antiquark (jets) 
as $q_1$ and $q_2$, respectively. They are on the mass shell, 
$q_1^2=q_2^2=0$.
We will use the Sudakov decomposition of 4-vectors with respect to 
the momenta of the incoming particles $p_1$ and $p_2$.
For instance, the jet momenta  are decomposed according to
\beq{sudakov}
q_1=zp_1+\frac{q_{1\perp}^2}{zs}p_2+q_{1\perp} \ ,
q_2=\bar zp_1+\frac{q_{2\perp}^2}{\bar zs}p_2+q_{2\perp}
\eeq
such that $z$ is the longitudinal momentum fraction of the quark 
jet in the lab frame.
We will often use the shorthand notation: $\bar u \equiv (1-u)$ for any
longitudinal momentum fraction $u$.
The Dirac spinors for the quark and the antiquark are denoted by   
$\bar u(q_1)$ and $v(q_2)$. 

We are interested in the forward limit, when the transferred momentum 
$t=(p_2-p_2^\prime)^2$ is equal to zero%
\footnote{If the target mass $m$ is taken into account, the momentum 
transfer $t=(p_2-p_2^\prime)^2$ contains a non-vanishing longitudinal 
contribution and is constrained from below by  
$|t|\geq t_0$, where $t_0=({\displaystyle m^2M^4})/{\displaystyle 
(s-m^2)^2}$, 
$M^2$ being the invariant mass of the dijet.}, 
and the transverse momenta of jets compensate each other
$q_{1\perp} \equiv q_{\perp}$, $q_{2\perp}\equiv -q_\perp$.
In this kinematics the invariant mass of the produced $q\bar q$ pair is
equal to 
$
M^2={\displaystyle q_{\perp}^2}/{\displaystyle z\bar z}
$.
The invariant c.m. energy $s=(p_1+p_2)^2=2p_1 p_2$ 
is taken to be much larger than the transverse jet momentum 
$q_{\perp}$. In what follows we neglect
contributions to the amplitude that are suppressed by powers of $1/s$.

The general scheme of the calculation can be explained as follows.
Since the hard scattering amplitude $T_H(z^\prime , x_1, \mu_F^2)$,
by assumption, does not depend on the target, we choose to 
consider the hard dijet production from a quark, $\pi q\to (\bar q q)q$.
We replace the pion by a collinear $\bar q q$ pair  with the momenta
$z'p_1$ and $\bar z'p_1$, respectively. The probability amplitude 
to find a particular value of the momentum fraction $z'$ is given by 
the pion distribution amplitude defined as
\beq{phi-pi}
\langle0|\bar d(y) \gamma_\mu \gamma_5u(-y)|\pi^+(p)\rangle_{y^2\to 0}
 =ip_{\mu}\, f_\pi
\int\limits^1_0 \!dz'\, e^{i(2z'-1)(py)} \phi_\pi (z')\,,
\eeq 
where $f_\pi\simeq 131$~MeV is the pion decay constant.
Light-cone dominance for the $t$-channel gluon emission 
is not assumed from the beginning, but has to follow from the
calculation in order that the result can be interpreted in the sense 
of the factorization formula \re{factor}. To this end, the gluon 
transverse momentum $k_\perp$ is kept nonzero and we show that
the amplitude in question contains a collinear logarithm 
$\ln q_{\perp}^2/\mu_F^2$ coming from the integration region 
$\mu_F^2 \ll k_\perp^2 \ll  q_{\perp}^2$. This property allows to   
calculate the upper part of the diagrams represented schematically 
in Fig.~\ref{fig:1} assuming that $k_\perp^2 \ll  q_{\perp}^2$ i.e.
in the light-cone limit. In the last step, the collinear logarithm
is interpreted as a contribution to the gluon distribution in 
the target and the (perturbative) non-forward gluon distribution 
of the quark is substituted by the (nonperturbative) non-forward 
gluon distribution in the nucleon. 

\begin{figure}[t]
\centerline{\epsfxsize12.0cm\epsffile{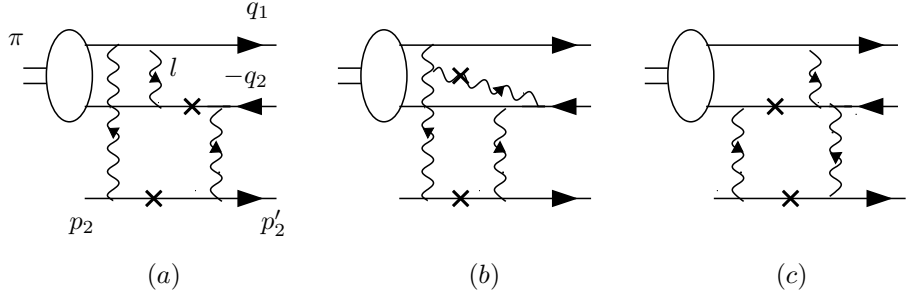}}
\caption[]{\small 
Cut diagrams (examples) for the imaginary part of the amplitude 
$\pi q\to (\bar q q)q$. The cut quark (gluon) propagators are indicated 
by crosses.  
 }
\label{fig:2}
\end{figure}

In this letter we calculate the 
imaginary part of the amplitude. The real part can in principle be 
restored from the imaginary part using dispersion relations. 
One rationale for this procedure is that  
at high energies the scattering amplitudes corresponding to 
Pomeron exchange are dominated by their imaginary parts.   
The second rationale is  simplicity:
the corresponding cut diagrams (see Fig.~\ref{fig:2} for examples) are 
built of tree-level on-shell scattering amplitudes and their form is 
strongly constrained by gauge invariance, see below.

%
%

\begin{figure}[htbp]
\centerline{\epsfxsize14.0cm\epsffile{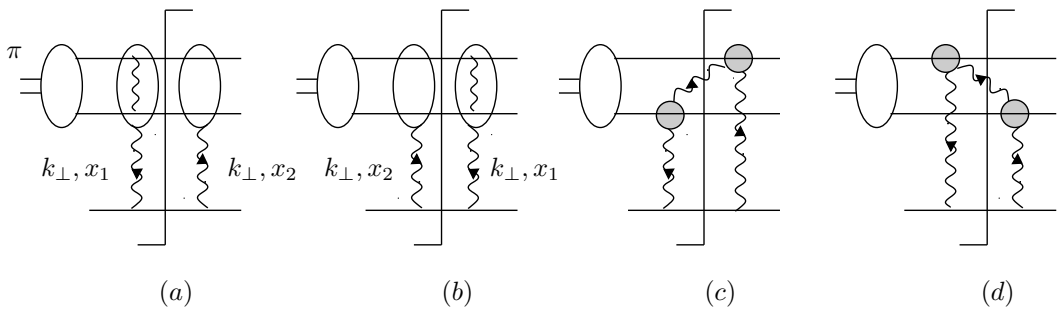}}
\caption[]{\small 
The decomposition of the imaginary part of the amplitude $\pi q\to (\bar q q)q$
into four gauge-invariant contributions.   
 }
\label{fig:3}
\end{figure}

\begin{figure}[t]
\centerline{\epsfxsize12.0cm\epsffile{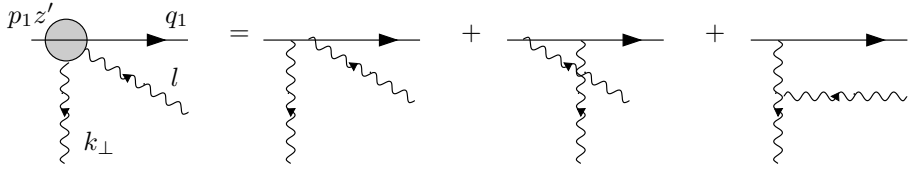}}
\caption[]{\small 
The effective vertex.  
 }
\label{fig:4}
\end{figure}

The existing cut diagrams can be  grouped into the  four 
gauge-invariant contributions shown in Fig.~\ref{fig:3}a--d,
%
%
which differ by the position of the hard gluon that provides the large 
momentum transfer to the jets. 
The corresponding contributions to the amplitude will be denoted 
as ${\cal M}_{(a)}$, ${\cal M}_{(b)}$, ${\cal M}_{(c)}$ and 
${\cal M}_{(d)}$.  For example, in  Fig.~\ref{fig:3}a it is 
assumed that the hard gluon exchange appears to the left of the cut.
This contribution is given by the sum of 10 Feynman diagrams one of which 
is shown in Fig.~\ref{fig:2}a. Similarly, the contribution in    
Fig.~\ref{fig:3}b is given by the sum of 10 diagrams with the hard gluon 
exchange appearing to the right of the cut; a typical diagram is shown 
in Fig.~\ref{fig:2}c. The two remaining contributions in Fig.~\ref{fig:3}c 
and Fig.~\ref{fig:3}d take into account the possibility of 
real gluon emission in the intermediate state. The filled circles 
stand for the effective vertices describing the gluon 
radiation, see Fig.~\ref{fig:4}. Each of the two contributions in   
Fig.~\ref{fig:3}c,d corresponds to a sum of 9 different Feynman diagrams, 
see Fig.~\ref{fig:2}b for an example.  

%
%

We use Feynman gauge and perform the usual substitution  
\begin{equation}
g_{\mu \,\nu} \rightarrow \frac{p_2^\mu\,p_1^\nu}{(p_1p_2)}
\label{6}
\end{equation}
in the propagators of $t-$channel gluons.  
This is correct up to terms  $\sim 1/s$ which is sufficient for our purposes. 
The $t-$channel gluon momenta are expanded according to
\bea{sudakov2}
k^{\mu}_1 &=& \alpha_1\,p_1^\mu + x_1\,p_2^\mu + k_\perp^\mu \,, 
\nonumber \\
k^{\mu}_2 &=& \alpha_2\,p_1^\mu + x_2\,p_2^\mu + k_\perp^\mu \,. 
\eea
As is easily checked by inspection, in any cut diagram 
two internal lines are on the mass shell.
The corresponding two on-shellness conditions fix $\alpha_1$ and $x_1$
and relate the variables $z^\prime$ and $x_1$ to one other. In addition,
$\alpha_2$ and $x_2$ are fixed by the energy--momentum conservation.
Since $\alpha_{(1,2)}$ and $x_{(1,2)}$ are all of the order of $1/s$,  
the $1/k_{1,2}^2$ factors in the propagators of the $t-$channel gluons 
can be approximated by 
$k_{(1,2)}^2=\alpha_{(1,2)} x_{(1,2)}s - k^2_{\perp}= - k^2_{\perp} +
{\cal O}(1/s)$. Using  the on-shellness conditions for the contributions
in Fig.~\ref{fig:3}a and Fig.~\ref{fig:3}b one obtains 
$x_1=\zeta$, $x_2=0$, for any $z^\prime $. 
For Fig.~\ref{fig:3}d one finds  
$x_1={\zeta z^\prime \bar z}/{(z^\prime -z)}$,
$x_2={\zeta z \bar z^\prime }/{(z^\prime -z)}$ and $z^\prime > z$, where
the last condition ensures that the energy of the cut gluon
is positive.
Finally, for the set of cut-diagrams corresponding to  
Fig.~\ref{fig:3}c  we obtain 
$x_1={\zeta z \bar z^\prime}/{(z-z^\prime)}$,
$x_2={\zeta z^\prime  \bar z}/{(z-z^\prime)}$ and $z > z^\prime$. 

After the on-shellness conditions are used, 
a single integration over the gluons transverse momentum $k_\perp$ remains:
\beq{impact-rep}
{\rm Im}\;{\cal M} \sim \int\;
\frac{d^2k_\perp}{(k^2_\perp
)^2}\;J_{up}(k_\perp,q_{\perp})\,J_{down}(k_\perp,q_{\perp})\,,
\eeq
where $k^4_\perp$ comes from the product of the two gluon propagators. 
$J_{up}$ and $J_{down}$ are dubbed impact factors and 
stand for the upper and the lower parts of the diagrams
Fig.~\ref{fig:2}a--d (connected by the two-gluon exchange). 
The representation \re{impact-rep} 
is  well known \cite{ChWu,LiFr} from QED
scattering at high energies.

Properties of the impact-factors $J_{up}$ and
$J_{down}$ as a functions of $k_{\perp}$ at $k_\perp\to 0$ 
are of crucial importance. 
Since $J_{down}$ is the impact-factor of a point-like target quark, 
$J_{down}(k_\perp,q_\perp)\sim const$. On the other hand,
$J_{up}(k_\perp,q_\perp)$ stands for the scattering of the colorless
$q\bar q$ (Fig.~\ref{fig:3}a--b) or $q\bar q G$ (Fig.~\ref{fig:3}c--d)
state having a transverse size $\sim 1/q_{\perp}$ and 
has to vanish at small $k_\perp \ll q_{\perp}$,
$J_{up}(k_\perp,q_\perp)\sim k^2_{\perp}$, as a consequence
of the color neutrality of the quark-antiquark pair: 
A gluon with a large wave length $\sim 1/k_\perp$ cannot resolve 
a color dipole of the small size $\sim 1/q_{\perp}$.
Since in our case there are two gluons, $J_{up}$ is proportional to 
the product $k_{\perp}\cdot k_{\perp}=k^2_{\perp}$%
\footnote{The ${\cal O}(k_\perp^2)$ behavior can be traced 
to the gauge invariance of the amplitude, see \cite{LiFr} for the details.}. 
In the opposite limit of large transferred momenta, 
$k_\perp\gg q_{\perp}$, the two 
$t-$channel gluons are forced to 
couple to the same parton (quark or gluon) in the 
upper block in Fig.~\ref{fig:3}a--d. It follows that at large $k_\perp$
$J_{up}(k_\perp,q_\perp)\sim const$.  

Taking into account the above properties of the impact-factors we 
conclude that the transverse momentum integration in \re{impact-rep}
diverges logarithmically at small $k_\perp$ and the integral can be 
estimated as ${\cal M} \sim \int^{q_\perp^2} dk_\perp^2/k_\perp^2 \sim 
\ln q_\perp^2$, as expected.    
The region of $k_\perp^2 > q_{\perp}^2$ 
does not produce the large logarithm and can be neglected.
Note that the correct small $k_\perp$ behavior of the impact factors 
is only recovered in the sum of cut diagrams for the gauge 
invariant amplitudes 
${\cal M}_{(a)}$, ${\cal M}_{(b)}$, ${\cal M}_{(c)}$ and 
${\cal M}_{(d)}$, but not for each diagram separately.   

In addition to the diagrams discussed so far, the amplitude 
$\pi q \to (\bar q q)q$ receives a contribution from the three-gluon 
exchange in the t-channel. Such terms can be viewed as belonging to the cut 
diagrams shown in Fig.~\ref{fig:3}a in which the hard gluon in the blob 
is attached  to the bottom quark line. We have checked that this extra  
contribution does not contain the large
collinear logarithm $\ln q_\perp^2$ and therefore we neglect it.

\vskip0.3cm
{\large \bf 3.~~} To start with, consider the calculation of ${\cal M}_{(d)}$.
Let $l^\mu=\alpha_lp_1^\mu+x_lp_2^\mu+l^\mu_\perp$ be the momentum of the 
(real) gluon in the intermediate state and let $e^\mu(l)$ be one
of the two physical polarization vectors.  The two conditions 
$(e\cdot p_2)=0$ and $(e\cdot l)=0$ fix the gauge and result in
$e^\mu(l) = e_\perp^\mu +
2p_2^\mu\,\left(e_\perp l_\perp\right)/(\alpha_l\,s)$.

The effective vertex corresponding to the sum of the three diagrams 
in  Fig.~\ref{fig:4} has the form
\beq{e-vertex}
i \frac{g^2\,z\,(z'-z)}
{q_{\perp}^2\, z'}\,
\left[\frac{1}{z'}\,\left(t^l\,t^a \right)_{ i \,j}-
\frac{1}{ z}\,\left(t^a\,t^l \right)_{ i \,j}  \right]
 {\bar u}(q_1)
\left[\!\not\! b\,\!\not\! e_\perp \ -2\,\frac{ z}{ z' -
z}(e_\perp b)\right]\,\frac{\!\not\! p_2}{s}\,u(z'p_1) \,. 
\eeq
Here $t^l$ and $t^a$ are the $SU(3)$ generators. The color indices $l$ and 
$a$ belong to the emitted gluon and the $t-$channel gluon,
respectively. We have also
introduced an auxiliary two-dimensional vector $b^\mu$ defined as:
\beq{b}
b^\mu = k_\perp^\mu - 2\,\frac{(k_\perp
q_{\perp})}{q_{\perp}^2}\,q_{\perp}^\mu\, ,\;\;\;\; b^2=k_\perp^2\,.
\eeq
Note that the effective vertex, in the limit of small
$k_\perp$,  is proportional to $b\propto k_\perp$. The constant terms
cancel in the gauge invariant sum of the diagrams in Fig.~\ref{fig:4}.

The second effective vertex in Fig.~\ref{fig:3}d has a similar form.
Combining both of them  and
performing the sum over the polarizations of the emitted gluon we obtain the 
impact-factor $J_{up}^{(d)}$.  
Since each effective vertex is proportional to $k_\perp$, it follows that 
$J_{up}^{(d)}\sim k^2_\perp$, as expected. The result for the amplitude 
${\cal M}_{(d)}$ is obtained using the representation in \re{impact-rep}.
The calculation of ${\cal M}_{(c)}$ is very similar. 
The result for their sum reads: 
\bea{Mcd}
\lefteqn{
{\cal M}_{(c)}+{\cal M}_{(d)} = D\,C_F^2\,\int\,\frac{d k_\perp^2}{k_\perp^2}
\,\int\limits_0^1dz'\,
\phi_\pi(z')\,\left(\frac{z\,\bar
z}{z'\,\bar z'}+1 \right)\times}
\nonumber \\
&\times&\left[\left(\frac{z\,\bar z}{z'\,\bar z'}+1\right)
+\frac{1}{(N_c^2-1)}\left(\frac{z}{z'}+ \frac{\bar z}{\bar z'}  \right)
\right]
 \left[ \frac{\Theta(z'-z)}{(z'-z)} + \frac{\Theta(z-z')}{(z-z')} \right],
\eea
where
\beq{D}
D= -i\,s\,f_\pi\,\alpha_s^3\,\frac{4\,\pi^2}{N_c^2\,q_{\perp}^4}\,{\bar
u}(q_1)\gamma_5 \frac{{\!\not\! p}_2}{s}v(q_2)\,
\delta_{i\,j}\,\delta_{c\,c'}\,,
\eeq
and $C_F= (N_c^2-1)/{2N_c}$. The color indices $(i,j)$ correspond to the 
produced quark-antiquark pair (jets) and $(c,c')$ stand for the color 
indices of the target quark in the initial and the final state.
The contributions $\sim \Theta(z'-z)$ and $\sim\Theta(z-z')$ 
belong to ${\cal M}_{(d)}$ and ${\cal M}_{(c)}$, respectively.

For the  cut diagrams in Fig.~\ref{fig:3}a and Fig.~\ref{fig:3}b 
we present the final results:
\bea{Mab}
{\cal M}_{(a)} &=& - D\,C_F^2\,\int\frac{d k_\perp^2}{k_\perp^2}
\,\int\limits_0^1 dz'\, \phi_\pi(z')
\left( \frac{\bar z}{z'}+ \frac{z}{\bar z'} \right),
\nonumber\\
{\cal M}_{(b)} &=& D\,C_F^2 \,\int\frac{d k_\perp^2}{k_\perp^2}
\,\int\limits_0^1 dz'\,
\frac{\phi_\pi(z')}{z'\bar z'}\,
\left[z \bar z\left(\frac{\bar z}{z'}+\frac{z}{\bar z'}
\right) +\frac{1}{(N^2-1)}\left(\frac{z\bar z}{z'\bar z'}+1
\right)   \right].
\eea

The transverse momentum integrals 
$\int dk_\perp^2/k_\perp^2 \sim \ln q_\perp^2$
in \re{Mcd} and \re{Mab} can be identified
with the (perturbative) non-forward gluon distributions of a quark:  
\beq{Fquark}
\frac{\alpha_s}{\pi}\, C_F\,\int\,\frac{d k_\perp^2}{k_\perp^2}=
\frac{\alpha_s}{\pi}\, C_F\,\ln q_{\perp}^2
\,\rightarrow \,\,{}_qF^g_{\zeta}(x).
\eeq
To justify this substitution, note that ${}_qF^g_{\zeta}(x)$ can be calculated 
to first order in perturbation theory from the evolution equation \cite{Rad96}%
\footnote{In our calculation $F_\zeta(x)$ enters 
in the DGLAP region $x\geq \zeta$ only, see \cite{Rad96} for more details.}
\bea{evolve}
  q_\perp^2\frac{d}{d q_\perp^2}\, {}_qF^g_{\zeta }(x,q_\perp^2)
 &=& \frac{\alpha_s}{2\pi}
\int\limits^1_x dz\,P^{gq}_\zeta (x,z)\,{}_qF^q_\zeta (z, q_\perp^2),
\nonumber\\
P^{gq}_{\zeta}(x,z)&=& C_F\left[
\left(1-\frac{x}{z}\right)\left(1-\frac{x-\zeta}{z-\zeta}\right)+1
\right].
\eea 
Using ${}_qF^q_\zeta (z) = \delta(1-z)$ and taking into account 
that in the high-energy region  $\zeta, x\ll 1$ the quark-gluon 
kernel simplifies to $P^{gq}_\zeta (x,1) = 2 C_F$,
we arrive at the substitution rule in Eq.~\re{Fquark}.
The final step is to replace ${}_qF^g_{\zeta}(x)$ by the (nonperturbative)
non-forward gluon distribution in the nucleon which is a nontrivial 
function of the parameters \cite{Rad96}:
\beq{Fxi}
\langle N(p^\prime )|y_\mu y_\nu G^a_{\mu\alpha}(0)  G^a_{\alpha\nu}(y)
|N(p)\rangle_{y^2\to 0}
 =\bar u(p^\prime)\!\not\! y u(p)\frac{(yp)}{2}
\int\limits^1_0\!dx\left[e^{-ix(py)}+e^{i(x-\zeta)(py)}
\right] F^g_\zeta (x) \,.
\eeq
Here $G^a_{\mu\nu}$ is the gluon field strength tensor. 
$u(p)$ and $\bar u(p^\prime)$ are the spinors of the initial 
and final nucleons. 

Our final result for the imaginary part of the amplitude for 
dijet production from a nucleon reads
\beq{ImM}
 {\rm Im}\,{\cal M} = 
  -i\,s\,f_\pi\,\alpha_s^2\,\frac{4\,\pi^3}{N_c^2\,q_{\perp}^4}\,{\bar
u}(q_1)\gamma_5 \frac{{\!\not\! p}_2}{s}v(q_2)\,{\cal I}\, \delta_{i\,j}\,
\eeq
with 
\bea{I}
{\cal I} &=& 
 \int\limits_0^1\!dz'\, \phi_\pi(z', \mu^2)
\left\{\left[C_F\left(\frac{z\bar z}{z'\bar z'}-1\right)
\left(\frac{\bar z}{z'}+\frac{z}{\bar z'}\right)
+\frac{1}{2N_c\,z' \bar z'}\left(\frac{z\bar z}{z'\bar z'}+1\right) 
\right]\,F_\zeta(\zeta,\mu^2) \right. 
\nonumber \\
&&\left. +\left(\frac{z\bar z}{z'\bar z'}+1  \right)
\left[C_F\left(\frac{z\bar z}{z'\bar z'}+1\right)+\frac{1}{2N_c}\left(\frac{z}{z'}
+\frac{\bar z}{\bar z'} \right)\right] \right. \nonumber \\
&&\left. \times\left[\frac{\Theta(z'-z)}{(z'-z)}\,F_\zeta\left(\frac{\zeta\,z'\bar
z}{z'-z},\mu^2
\right) + \frac{\Theta(z-z')}{(z-z')}\,F_\zeta\left(\frac{\zeta\,\bar z'
z}{z-z'},\mu^2
  \right)    \right] \right\}
\eea
The differential cross section summed over the polarizations and the color
of quark jets is given by 
\beq{cross}
\frac{d\sigma_{\pi\to 2\,{\rm jets}}}{d q_\perp^2 dt dz} 
= \frac{\alpha_s^4 f_\pi^2 \pi^3}{8N_c^3 q_\perp^8} |{\cal I}|^2\,.
\label{crosssection}
\end{equation}
The factorization scale $\mu^2$ has to be 
of order of the transverse momentum of the exchanged gluon.
The expression in Eq.~\re{I} presents the main result of this paper.

\vskip0.3cm
{\large \bf 4.~~}The integrand in \re{I} is singular at $z'=z$, i.e. 
when the longitudinal momentum fraction carried by the quark coincides 
with that of the quark jet in the final state, and at the end-points 
of the integration region $z'\to 0$ and $z'\to 1$. Let us discuss the 
contributions from these regions in some detail.

The singularity at $z'=z$ is present in the contributions in 
Fig.~\ref{fig:3}c,d which include real gluon emission in the intermediate
state. The logarithmic integral $\int dz'/|z-z'| \sim \ln s$ is nothing but 
the usual energy logarithm that accompanies each extra gluon in the 
gluon ladder. Its appearance is due to the fact the the gluon 
in Fig.~\ref{fig:3}c,d can be emitted in a broad rapidity interval and
is not constrained to the pion fragmentation region.
To logarithmic accuracy we can simplify the integrand in \re{I}
by assuming $z'=z$ everywhere except for the diverging denominators and the 
argument of the gluon distribution, to get 
\beq{z=z'}
{\cal I}\Big|_{z'\approx z}
= 4N_c \,\phi_\pi(z)\,\int\limits^{1}_{z}
  \frac{dz'}{z'-z} F_\zeta(\zeta\frac{z'\bar z}{z'-z},q^2_{\perp})
\simeq 4N_c \,\phi_\pi(z)\!\int\limits_\zeta^1 
 \!\frac{dy}{y}\, F_\zeta(y,q^2_{\perp})\,.
\eeq
For a flat gluon distribution $F_\zeta(y) \sim {\rm const}$ at $y\to 0$, 
and the integration gives $ {\rm const}\cdot \ln 1/\zeta $ which is the above
mentioned logarithm. 
Note that the color factors combine to produce $C_A =N_c$ signaling that 
the relevant Feynman diagrams in Fig.~\ref{fig:3}c,d are those with a 
three-gluon coupling. Moreover, the factor $2N_c/y$ appearing in \re{z=z'}
can be interpreted as the relevant limit of the DGLAP splitting function
\cite{Rad96}
\beq{NNN}
q_\perp^2 \frac{\partial}{\partial q_\perp^2}\,F_\zeta(x=\zeta,q_\perp^2)
 = \frac{\alpha_s}{2\pi} \int\limits_{\zeta}^1 dy\,P_\zeta^{gg}(\zeta,y)\,
F_\zeta(y,q_\perp^2)
 \simeq \frac{\alpha_s}{2\pi} \int\limits_{\zeta}^1 dy\,\frac{2N_c}{y}\,
F_\zeta(y,q_\perp^2)\,.
\eeq 
The quantity on the l.h.s. of \re{NNN} defines what can be called the 
unintegrated non-forward gluon distribution and the physical meaning of 
Eqs.~\re{z=z'} and \re{NNN} is that in the region $z'\sim z$ hard 
gluon exchange can be viewed as a large transverse momentum part of the 
gluon distribution in the proton, cf. \cite{NSS99}.
This contribution is proportional to 
the pion distribution amplitude $\phi_\pi (z,q^2_{\perp})$ 
and contains the enhancement factor $\ln 1/\zeta \sim \ln s/q_\perp^2$. 

Next, consider the contribution to the imaginary part of the amplitude 
of the dijet production coming from the end-points $z'\to 0$ and 
$z'\to 1$. Summing both of them and using the symmetry of the 
pion distribution amplitude we obtain 
\beq{end}
 {\cal  I}\Big|_{\rm end-points}
 =  \left(N_c+\frac{1}{N_c}\right)
 z \bar z \,  \int\limits^1_0 dz'\,
\frac{\phi_\pi(z',\mu^2)}{\bar z'^2}F_\zeta(\zeta,\mu^2) \,.
\eeq
Since $\phi_\pi(z')\sim z'$ at $z'\to 0$, the integral over $z'$ 
diverges logarithmically. This divergence indicates that the 
collinear factorization conjectured in \re{factor} is generally not valid.
Remarkably, the divergent integral containing the pion distribution amplitude
is just a constant and does not involve any $z$-dependence.
Therefore, the longitudinal momentum distribution of the jets in the 
nonfactorizable contribution is calculable and, as it turns out,
has the shape of the asymptotic pion distribution amplitude 
$\phi_\pi^{\rm as}(z) = 6z\bar z$. The corresponding physical process 
is the following. The limit $z'\to 1$
corresponds to a kinematics in which the quark carries the entire
momentum of the pion. The fast quark radiates a hard gluon 
which carries the fraction $(1-z)$ of quark momentum. 
This radiation is perturbative and  
is described by the effective vertex \re{e-vertex} at $z'=1$.
At the final step the hard gluon transfers its entire longitudinal and 
transverse momentum to the quark jet, and emits a soft antiquark which 
interacts nonperturbatively with the target proton 
and the pion remnant. 
The technical reason for the singularity at  $z'\to1$  
can be traced to the quark propagator whose denominator has the form 
$ \sim 1/[
q^2_\perp {\bar z}^{'2}/({\bar z}(\bar z -\bar z')) - 2{\bar z'}(k_\perp
q_\perp)/(\bar z -
\bar z') + k_\perp^2 (1+\bar z'/(\bar z - \bar z'))
+ i\epsilon]$,
where $k_\perp$ is the transverse momentum of the $t$-channel gluon.
In order to extract the leading small-$k_\perp$ behavior corresponding to the
logarithmic collinear divergence we expand this denominator 
at small $k_\perp$. The leading term in this expansion 
cancels in the gauge invariant sum of diagrams, and the second term 
(linear in $k_\perp$ ) plus similar terms from the nominator 
produce $1/(1- z')^2$. It follows that the 
divergent logarithm in $\int \phi_\pi(z')/\bar z'^2$ is of the form 
$\ln q_\perp^2/\mu_{\rm IR}^2$ where $\mu_{\rm IR}$ is related to the   
average transverse momentum of the quarks inside the pion. It is possible that 
in the case of scattering from a heavy nucleus $\mu_{\rm IR}$ may grow as 
$\sim A^{1/3}$ because of color filtering. 
A detailed discussion of this effect goes beyond the tasks of this letter.

\vskip0.3cm
{\large \bf 5.~~}In order to make an estimate we assume that 
skewedness does not affect significantly the $x$ dependence and adopt
 the following simple model for the non-forward gluon distribution:
\beq{model1}
{}F_\zeta(x,\mu) = F_\zeta(\zeta,\mu)\,\left(\frac{\zeta}{x}\right)^\Delta\,
\eeq 
with $\Delta =0.3$ \cite{FFS98}. We further notice that at high energies, 
or at small $\zeta$, there exists a mechanism
which effectively suppresses the nonperturbative end-point contributions,
or, better to say, enhances the factorizable contributions in comparison 
to the end-point ones.
It is known that in the region of small $x$ the scale dependence
of the gluon density $g(x,\mu^2)$ is quite large and can be parametrized
by the effective exponent $\gamma$:
$g(x,\mu^2)=g(x,\mu^2_0)({\mu^2}/{\mu^2_0})^\gamma$.
We take $\gamma =0.3$ \cite{MRT97} and assume the same scale 
dependence for the non-forward distribution:
\beq{model2}
F_\zeta(\zeta,\mu^2)=F_\zeta(\zeta,\mu^2_0)\left( \frac{\mu^2}{\mu^2_0}
\right)^\gamma\,.
\eeq
It follows that the contributions coming from larger scales (alias smaller
transverse distances) are enhanced in comparison with the contributions 
coming from smaller scales (larger transverse distances). The typical
virtualities of the quark and gluon propagators in the hard
subprocess are $\sim q^2_\perp$ when $z'\sim z$. They decrease 
when $z'$ is close to the end points: $\sim (z'/z)q^2_\perp$ at
$z'\to 0$ and $\sim (\bar z'/\bar z)q^2_\perp$ at $z'\to 1$, as
the reminder that the soft contribution originates from large transverse
distances.
Therefore, it is natural to choose as the factorization scale:
\beq{scale}
\mu^2=q_\perp^2\,\frac{z'\bar z'}{z\bar z} \,.
\eeq
With this choice we observe that 
the end-point singularities (formally) disappear.

\begin{figure}[t]
\centerline{\epsfxsize9.0cm\epsffile{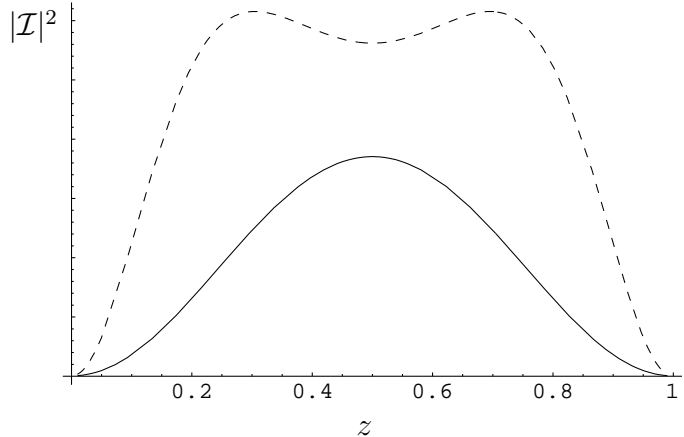}}
\caption[]{\small 
The longitudinal momentum fraction dependence of the jets
for two different choices of the 
pion distribution amplitude: $\phi^{\rm as}_\pi(z) = 6 z\bar z$ and
$\phi^{\rm CZ}_\pi(z) = 30 z\bar z (2z-1)^2$ shown by the solid and the 
dashed curve, respectively.
 }
\label{fig:5}
\end{figure}

The longitudinal momentum fraction dependence of the jets  
calculated using the expression in \re{I} and the model 
for the nonforward gluon distribution in \re{model1}, \re{model2} 
is shown in Fig.~\ref{fig:5} for two different choices of the 
pion distribution amplitude: $\phi^{\rm as}_\pi(z) = 6 z\bar z$ and
$\phi^{\rm CZ}_\pi(z) = 30 z\bar z (2z-1)^2$ \cite{CZ}.
For this plot we have taken $s=1000$~GeV$^2$ and $q_\perp = 2$~GeV
which roughly corresponds to the kinematics of the 
E791 experiment \cite{E791}. The overall normalization is arbitrary.
Notice that the Chernyak-Zhitnitsky (CZ) ansatz for the pion distribution 
amplitude leads to a much larger (integrated) cross section than the 
asymptotic distribution\footnote{The calculation shown in Fig.~\ref{fig:5}
serves the illustrative purposes only. In the data analysis
one has to take into account that the CZ model \cite{CZ} 
is formulated at a low scale and has to be evolved to $\mu \sim q_\perp$ for 
a meaningful comparison. }. 
 This  signals that the leading logarithmic 
approximation in Eq.~\re{z=z'} is not sufficient and regions of 
integration in \re{I} other than $z\approx z'$ play an important r\^ole. 
%
%
One sees that the sensitivity to the shape of the pion distribution
amplitude remains, although there is no direct proportionality
as it was assumed in \cite{E791}.

To summarize, in this letter we have calculated the hard gluon exchange 
contribution to the imaginary part of the amplitude of pion diffraction 
dissociation to two jets with large transverse momenta. 
The answer is given in Eqs.~\re{I}, \re{cross} and its main feature is that
the jet longitudinal momentum distribution is not simply proportional 
to the pion distribution amplitude. The actual dependence 
is rather elaborate and it has to be taken into account in the 
data analysis. 
Our result can be improved by calculating the quark contribution
which may be important in the energy range of the E791 experiment and 
by elaborating on possible nuclear effects that were not taken 
into account in the present study.

\vspace*{0.3cm}

{\Large\bf Acknowledgments}

\vspace*{0.3cm}

\noindent V.B. is  grateful to A. Radyushkin for a discussion and critical
remarks. L.Sz. and D.I. were supported by the DFG 
and the Alexander von Humboldt Stiftung, respectively. V.B. acknowledges
warm hospitality at the Instutute of Nuclear Theory where this work was
finalized. 

\vspace*{0.3cm}

{\Large\bf Note added}

\vspace*{0.3cm}

\noindent
Our result in Eq.~(\ref{I}) does not agree with an independent calculation 
\cite{Chernyak:2001ph} using a different method where the light-cone 
dominance was assumed from the beginning. We plan to investigate the 
reasons for this disagreement in a separate publication.


\begin{thebibliography}{99}

\bibitem{BBGG81}
G.~Bertsch, S.~J.~Brodsky, A.~S.~Goldhaber and J.~F.~Gunion,
Phys.\ Rev.\ Lett.\ {\bf 47} (1981) 297.

\bibitem{FMS}
L.~Frankfurt, G.~A.~Miller and M.~Strikman,
Phys.\ Lett.\ B {\bf 304} (1993) 1;
%
Found.\ Phys.\ {\bf 30} (2000) 533 [hep-ph/9907214];
%
hep-ph/0010297.

\bibitem{NSS99}
N.~N.~Nikolaev, W.~Schafer and G.~Schwiete,
Phys.\ Rev.\ D {\bf 63} (2001) 014020.

\bibitem{E791}
E.~M.~Aitala {\it et al.}  [E791 Collaboration],
hep-ex/0010043.

\bibitem{Rad96}
A.~V.~Radyushkin,
Phys.\ Lett.\ B {\bf 385} (1996) 333;
Phys.\ Rev.\ D {\bf 56} (1997) 5524.

\bibitem{Ji97}
X.~Ji,
Phys.\ Rev.\ Lett.\ {\bf 78} (1997) 610;
J.\ Phys.\ G{\bf G24} (1998) 1181.

\bibitem{ChWu} H. Cheng and T.T. Wu, Phys.\ Rev.\ 182 (1969) 1852.

\bibitem{LiFr} L.N. Lipatov and G.W. Frolov, Sov. Yad. Fiz. 13, (1971) 588.


\bibitem{asymptotic}
A.~V.~Efremov and A.~V.~Radyushkin,
Phys.\ Lett.\ B {\bf 94} (1980) 245;
Theor.\ Math.\ Phys.\ {\bf 42} (1980) 97;\\
G.~P.~Lepage and S.~J.~Brodsky,
Phys.\ Lett.\ B {\bf 87} (1979) 359;
Phys.\ Rev.\ D {\bf 22} (1980) 2157;\\
S.~J.~Brodsky, Y.~Frishman, G.~P.~Lepage and C.~Sachrajda,
Phys.\ Lett.\ B {\bf 91} (1980) 239.

\bibitem{CZ}
V.~L.~Chernyak and A.~R.~Zhitnitsky,
Nucl.\ Phys.\ B {\bf 201} (1982) 492.
Phys.\ Rept.\ {\bf 112} (1984) 173.

\bibitem{FFS98}
L.~L.~Frankfurt, A.~Freund and M.~Strikman,
Phys.\ Rev.\ D {\bf 58} (1998) 114001.

\bibitem{MRT97}
A.~D.~Martin, M.~G.~Ryskin and T.~Teubner,
Phys.\ Rev.\ D {\bf 55} (1997) 4329.


\bibitem{Chernyak:2001ph}
V.~Chernyak,
hep-ph/0103295.

\end{thebibliography}
\end{document}